\documentclass[preprint,12pt]{elsart}
\usepackage{graphicx,amssymb,amsmath,times}
\usepackage{setspace}
\journal{Astroparticle Physics}

\begin{document}
\begin{frontmatter}
\title{Study of short term enhanced TeV $\gamma$- ray emission from Mrk 421 observed with TACTIC \\
        on December 28, 2014}
\author[1]{K K Singh\corauthref{cor}},
\corauth[cor]{Corresponding author.}
\ead{kksastro@barc.gov.in}
\author[1,2]{K K Yadav}, \author[1]{K Chanchalani}, \author[1]{P Chandra}, \author[1]{B Ghosal}, \author[1,2]{A K Tickoo}, \author[1,2]{R C Rannot}, 
\author[1]{P Marandi}, \author[1]{N K Agarwal}, \author[1]{M Kothari}, \author[1]{K K Gour}, \author[1]{H C Goyal}, \author[1]{A Goyal}, \author[1]{N Kumar}, 
\author[1]{C Borwankar}, \author[1]{S R Kaul}, \author[1]{V K Dhar}, \author[1]{M K Koul}, \author[1]{R Koul}, \author[1]{K Venugopal}, \author[1]{C K Bhat}, 
\author[1]{N Chouhan}
\address[1]{Astrophysical Sciences Division, Bhabha Atomic Research Centre, Mumbai- 400 085, India} 
\address[2]{Homi Bhabha National Institute, Mumbai - 400 094, India}

\begin{abstract}
In this work, we report on the detection of enhanced TeV $\gamma$- ray emission from the high synchrotron-peaked blazar Mrk 421 with the TACTIC 
telescope on the night of December 28, 2014 (MJD 57019). We use data from the TACTIC observations of Mrk 421 for one week during December 25-31, 
2014 (MJD 57016-57022) in this study. The TACTIC observation on December 28, 2014 (MJD 57019) alone results in the detection of 86$\pm$17 $\gamma$- ray 
like events from Mrk 421 with a statistical significance of 5.17$\sigma$ in a livetime of $\sim$ 2.2 hours above an energy threshold of 0.85 TeV. 
The high statistics (higher than three Crab Units) of TeV photons enables us to study the very high energy (VHE) $\gamma$- ray emission from the source at
shorter timescales. A minimum variability timescale of $\sim$ 0.72 days is obtained for the TeV $\gamma$- ray emission from Mrk 421 during the above flaring 
activity of the source. The intrinsic VHE spectrum is described by a power law with spectral index of 2.99$\pm$0.38 in the energy range 0.85--8.5 TeV. 
The integral VHE $\gamma$- ray flux above 0.85 TeV is determined to be (3.68$\pm$0.64)$\times$10$^{-11}$ ph~cm$^{-2}$~s$^{-1}$ from the TACTIC observations of 
Mrk 421 on the night of December 28, 2014 (MJD 57019). Near simultaneous measurements by the HAWC observatory give an integral flux of 
(2.90$\pm$0.40)$\times$10$^{-11}$ ph~cm$^{-2}$~s$^{-1}$ above 2 TeV from Mrk 421 observations on December 29, 2014 (MJD 57020.33-57020.58). We have also analyzed 
the contemporaneous data from \emph{Fermi}-LAT to study the high energy (HE) $\gamma$--ray emission during the high activity state of the source. The HE $\gamma$--ray 
emission is observed to be increasing after the TeV flaring activity detected with the TACTIC. We also use other near simultaneous archival data available from the 
\emph{Swift}-BAT in hard X-rays and from SPOL at Steward Observatory in optical V and R bands to characterize the multi-wavelength emission of Mrk 421 during the high activity 
state observed at TeV energies. The TeV $\gamma$- ray emission observed on December 28, 2014 (MJD 57019) is found to be more prominant than the emissions in lower energy 
bands during the same period. The TeV $\gamma$- ray observation of Mrk 421 in high activity state with the TACTIC telescope is also used to understand the physical mechanism for 
blazar emission under the frame work of the leptonic single zone synchrotron self Compton process.    
\end{abstract}

\end{frontmatter}

\section{Introduction}  
Mrk 421 is a relatively nearby blazar located at a distance of $\sim$ 135 Mpc (redshift z=0.031) in the extragalactic sky [1]. It has been classified as 
the high synchrotron-peaked (HSP) blazar on the basis of the position of synchrotron peak frequency ($\nu_{syn}^p \ge 10^{15}$Hz) in the spectral energy 
distribution (SED) of blazars, which is generally characterized by a double hump structure [2]. Mrk 421 is one of the well studied and strongest 
TeV $\gamma$- ray sources in the northern hemisphere. Motivated with the detection of high energy (HE) $\gamma$- ray emission from Mrk 421 above 100 MeV 
by \emph{EGRET} for the first time in 1991 [3], the source was selected as a prime TeV candidate for very high energy (VHE) $\gamma$- ray observation with 
the ground based Whipple telescope. In March-June 1992, the Whipple telescope discovered the first unambiguous VHE $\gamma$- ray emission from Mrk 421 with 
a statistical significance of 6$\sigma$ above 0.5 TeV and an integral flux of 30$\%$ of the Crab Nebula flux [4]. After Whipple observations, Mrk 421 
became the first extragalactic source detected at TeV energies. Since its discovery at TeV energies, Mrk 421 has been observed to exhibit episodes of 
strong flaring activities over the entire electromagnetic spectrum from TeV $\gamma$- rays to radio energies on several occasions. 
During 1993-2004, the Whipple telescope detected many dramatic outbursts at TeV energies from Mrk 421 with doubling timescales from hours to less 
than 15 minutes [5,6] and for the first time spectral hardening of TeV $\gamma$- ray emission during the flare was observed in a blazar [7,8]. 
The MAGIC telescope observed TeV $\gamma$- ray emission from this source in 2004 [9] followed by first simultaneous observation of X-ray and 
TeV flares in 2006 [10]. Subsequently, this blazar had been regularly monitored by all the ground based $\gamma$- ray telescopes with the 
detection of various flaring activities from the source [11,12,13]. Again, a major outburst at all energies was observed from Mrk 421 in February 2010 
by various ground and space based instruments. At TeV energies, this flaring activity was detected by VERITAS, HESS, TACTIC, HAGAR and ARGO-YBJ detectors 
[14,15,16,17,18,19,20]. Apart from the detection of multiple short-term flaring activities from Mrk 421 during the last two decades, many long term 
multi-wavelength observations of the source are also reported in the literature by various telescopes including TACTIC [21,22,23,24,25,26,27,28,29,30]. 
An integral baseline flux of 33$\%$ of the Crab Nebula flux above 1 TeV has been derived for Mrk 421 using a combination of data collected during 1991-2009 [31].
\par
Blazars are radio-loud active galactic nuclei (AGNs) powered by accretion on to the supermassive black holes in the Universe. They are characterized by the 
relativistic jets originating from the region close to the central engine and pointed towards the line of sight from the Earth. The relativistic effects like 
Doppler boosting of the non-thermal radiation emitted from the blazar-jet are more pronounced and the emission is observed to be variable over the entire 
eletromagnetic spectrum from radio to VHE $\gamma$- rays. Mrk 421 has also been an important blazar to investigate correlations in TeV $\gamma$- ray and X-ray 
fluxes measured during low and high activity states. A tentative positive and strong correlation between X-ray and TeV $\gamma$- ray fluxes is found during 
multi-wavelength campaign of several flaring episodes [32,33]. However, VHE $\gamma$- ray flares without any X-ray activity have also been observed [8,34]. 
Such flaring activities are referred to as \emph{orphan} TeV flares. A positive correlation in X-ray and TeV $\gamma$- ray emission during quiescent state of 
Mrk 421 is also reported [35]. The connection between the variations in TeV energy bands and lower energy bands has not been clearly understood for blazars 
like Mrk 421 and detailed time-dependent emission models are being developed to study the complex correlations among different energy bands [36,37,38].
\par
Because of its proximity and high degree of multi-wavelength variability at different timescales, Mrk 421 has been a good extragalactic TeV source for 
understanding the physical mechanisms involved in the blazar emission during the quiescent as well as flaring states. The low energy emission from blazars 
is attributed to the relativistically beamed incoherent synchrotron radiation whereas high energy emission in GeV-TeV regime has not yet been well understood. 
Different models have been proposed in the literature to explain the $\gamma$- ray emission from blazars in quiescent as well as flaring states. In the leptonic 
synchrotron self Compton (SSC) model, the high energy $\gamma$- ray photons are produced by the inverse Compton (IC) scattering  of the low energy synchrotron 
photons by the same population of relativistic electrons that emit the synchrotron photons [39,40,41]. In another leptonic model, the target photons for IC enter 
from outside regions like accretion disk [42], broad-line region and dusty torus [43]. This is referred to as the External Compton (EC) model for $\gamma$- ray 
emission in blazars. On the other hand, hadronic models have also been proposed in which $\gamma$- ray photons are produced by proton synchrotron emission [44,45] 
and by the secondary particles of the proton-initiated cascades [46,47].     
\par
Motivated by the observation of frequent flaring activities of Mrk 421, we study the sudden increase in the TeV $\gamma$- ray emission on the night of 
December 28, 2014 (MJD 57019) observed with the TACTIC. In order to characterize the short term enhanced TeV $\gamma$- ray emission from the source, we have used 
data from TACTIC observations of Mrk 421 collected during December 25-31, 2014 including the high activity state. In Section 2, we describe the 
observations and data analysis procedures followed in different energy bands for the period December 25-31, 2014 (MJD 57016-57022). The results from the TACTIC observations 
on the night of December 28, 2014 (MJD 57019) are presented in Section 3. In Section 4, results from broad-band near simultaneous observations during 
December 25-31, 2014 in multi-wavelength context are reported. Finally, we conclude our study in Section 5. We have adopted $\Lambda$CDM cosmology 
with parameters H$_0$ = 70 km s$^{-1}$ Mpc$^{-1}$, $\Omega_m$ = 0.27 and $\Omega_{\Lambda} $= 0.73 throughout this paper.
\section{Observations and data analysis}

\subsection{TACTIC: VHE $\gamma$- rays }
The TeV Atmospheric Cherenkov Telescope with Imaging Camera (TACTIC) is a VHE $\gamma$- ray telescope located at GOALS (Gurushikhar Observatory for 
AstrophysicaL Sciences) Observatory (24.6$^\circ$ N, 72.7$^\circ$ E, 1.3 km asl), Mount Abu, India [48]. The telescope deploys a F/1-type tesselated light 
collector of $\sim$9.5 m$^2$ area, with a 349-pixel photomultiplier based imaging camera at its focal plane. 
The TACTIC telescope has undergone a major upgrade in 2011 to improve its over all performance. With its current hardware configuration the TACTIC telescope can detect a VHE 
$\gamma$- ray signal above an energy threshold of 0.85 TeV from the Crab Nebula like source at a statistical significance of 5$\sigma$ in 12 hours of observation time. 
The telescope has an angular resolution of $\sim$ 0.22$^\circ$ and an energy resolution of $\sim$ 26$\%$ at 1 TeV. The VHE $\gamma$- ray data used in this study was collected from the 
Mrk 421 direction with the TACTIC telescope during December 25-31, 2014 (MJD 57016-57022). We have applied standard data quality checks (compatibility of the prompt coincidence rate 
with the expected zenith angle trend, Poisson distribution for the arrival times of the cosmic ray events and steady behaviour of the chance coincidence rates with time) to the raw data 
collected with TACTIC for obtaining the clean data for further analysis. The application of data quality checks results in a live observation time of $\sim$ 11 hours. In the next step, 
the clean data are analysed using the analysis procedure developed for the TACTIC telescope. We have followed the standard analysis procedure based on Hillas parameter technique 
(where each extensive air shower image recorded in the telescope camera is characterized by its moments of various order) [49,50] to separate the $\gamma$- ray like events from the huge 
background of hadronic events. The energy of $\gamma$- ray like events is estimated using an artificial neural network based methodology developed for the TACTIC telescope [51]. Details of 
the data analysis procedure used for TACTIC after upgrade can be found in [30,52,53].

\subsection{\emph{Fermi}-LAT: HE $\gamma$- rays }
The Large Area Telescope (LAT) on board  the \emph{Fermi} satellite is a pair-conversion HE $\gamma$- ray telescope optimized for exploring the sky in the energy range 
from 30 MeV to beyond 300 GeV [54]. The \emph{Fermi}-LAT scans the entire sky approximately every three hours in survey mode with a wide field of view and 
an unprecedented sensitivity for the detection of HE $\gamma$- ray photons. We have analysed the publically available Pass 8 data\footnote{https://fermi.gsfc.nasa.gov/cgi-bin/ssc/LAT/LATDataQuery.cgi} 
for the period December 25-31, 2014 (MJD 57016-57022) using the \emph{Fermi} ScienceTools software package version \emph{v10r0p5}. We have followed the standard procedure 
for the unbinned likelihood analysis of the LAT data in the energy range 100 MeV--300 GeV for a region of interest (ROI) of 10$^\circ$ radius centered at the position of Mrk 421. 
A maximum zenith angle cut of 90$^\circ$ is applied for $SOURCE$ events to reduce contamination from the Earth's limb where $\gamma$--rays are expected to be produced by interactions 
of cosmic-rays with the upper atmosphere of the Earth. The instrument response function P8R2$\_$SOURCE$\_$V6 with the diffuse $\gamma$--ray emission model files gll$\_$iem$\_$vo6.fit and 
iso$\_$P8R2$\_$SOURCE$\_$V6$\_$v06.txt for Galactic and extragalactic components respectively are used in the analysis. All the point sources within 20$^\circ$ from Mrk 421 have 
been included in the model file from the third \emph{Fermi} $\gamma$--ray source catalog (3FGL) [55]. The significance of HE $\gamma$--ray events from the source under 
study is estimated using likelihood ratio test statistic (TS) defined in [56]. The spectral parameters of all the sources in the optimized model file are fixed to their 
catalog values except Mrk 421 for which the spectral parameters of a power law model are left free for generating the daily light curve. We have computed upper limits on the integral 
flux at 2$\sigma$ confidence level for the light curves with TS $\le$ 25. For the estimation of spectral flux points in different energy bins, the spectral index of the source is fixed 
to the value obtained from the best-fit over a given time period for which spectral analysis is performed and the normalization is left free to vary. For the background emission models,
the normalization is left free during the calculation of light curves and flux points for the spectra of the source.

\subsection{\emph{Swift}-BAT: Hard X-rays}
The Burst Alert Telescope (BAT) on board the \emph{Swift} satellite is a coded aperture mask telescope operating in the hard X-ray energy range 14-195 keV [57]. 
The \emph{Swift}-BAT monitors the whole sky every 1.5 hours while orbiting the Earth. The daily light curves from the sources detected by \emph{Swift}-BAT are 
provided online\footnote{https://swift.gsfc.nasa.gov/results/transients} in the energy range 15-50 keV [58]. We have used the archival data available from Mrk 421 
during the period December 25-31, 2014 (MJD 57016-57022) to obtain the daily light curves of the source. The photon flux measurements have been converted into the energy flux values 
using the appropriate mean energy of the hard X-ray photons in the energy range 15-50 keV.
 
\subsection{SPOL: Optical}
The Spectro-POLarimeter (SPOL) at \emph{Steward Observatory} of the University of Arizona provides optical observations for the LAT-monitored blazars under \emph{Fermi} multi-wavelength 
blazar monitoring program [59]. We have used the public archival data\footnote{http://james.as.arizona.edu/~psmith/Fermi/DATA/Objects} for Mrk 421 observations in R and V bands 
available during December 25-31, 2014 (MJD 57016-57022). The observed magnitudes in R and V bands have been converted into corresponding energy flux values using appropriate zero magnitude 
flux points to build the daily optical light curves of the source.

\begin{figure}
\begin{center}
\includegraphics[width=1.0\textwidth]{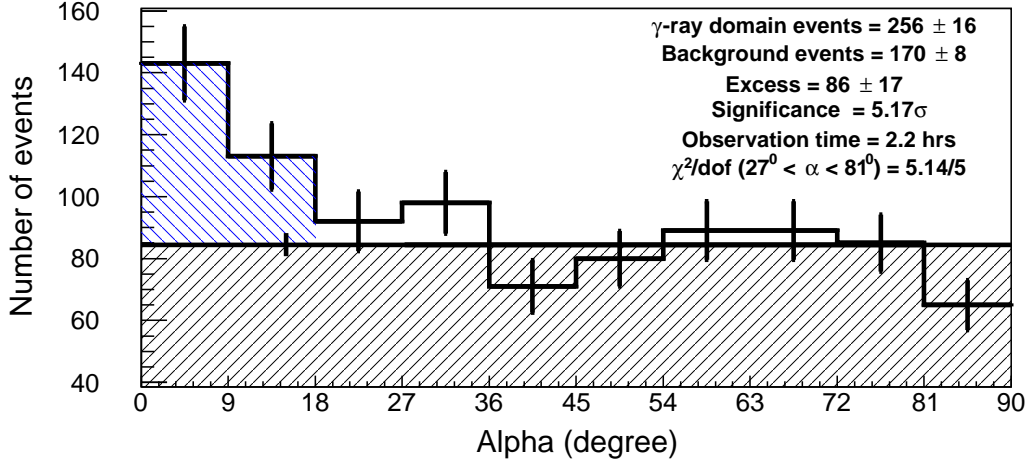}
\caption{Distribution of $\alpha$-parameter for Mrk 421 observation with the TACTIC telescope on the night of December 28, 2014 (MJD 57019). The blue shaded region indicates 
         the $\gamma$- ray domain ($0^\circ \le \alpha \le 18^\circ$) from the source direction and the black shaded region represents the level of isotropic cosmic-ray hadronic 
         background ($27^\circ \le \alpha \le 81^\circ$).}
\label{fig:Fig1}
\end{center}
\end{figure}
\section{Results from TeV observations on December 28, 2014 with TACTIC}

\subsection{TeV $\gamma$- ray detection}
In VHE $\gamma$- ray observations with ground based imaging Cherenkov telescopes, detection of the $\gamma$- ray signal from a source direction is determined from the frequency 
distribution of $\alpha$- parameter which is defined as the angle between the major axis of the shower image and the line connecting the camera center to the image centeroid. 
The frequency distribution of $\alpha$- parameter is expected to be flat for the isotropic cosmic-ray hadronic events, whereas it shows a peak at smaller $\alpha$- values 
($\le$ 18$^\circ$) for $\gamma$- ray events coming from a point source. The statistical significance of the detected $\gamma$- ray like events is determined using the methodology 
proposed by Li and Ma (1983) [60]. The distribution of $\alpha$- parameter for the events detected during the observation of Mrk 421 with the TACTIC on the night of December 28, 
2014 (MJD 57019.89583-57019.97917) is shown in Figure \ref{fig:Fig1}. The livetime of $\sim$ 2.2 hours from Mrk 421 observation with TACTIC on December 28, 2014 has resulted in the 
detection of 86$\pm$17 VHE $\gamma$- ray photons with a statistical significance of 5.17$\sigma$. The time averaged event rate from Mrk 421 during the above period is found to be 
(39$\pm$8) $\gamma$- rays per hour, which corresponds to $\sim$ 3- times the $\gamma$- ray rate from the Crab Nebula (for TACTIC, Crab Unit corresponds to the detection of TeV photons 
at an average rate of $\sim$ 14 $\gamma$- rays per hour or an integral flux of $\sim$ 1.12$\times$ 10$^{-11}$ ph~cm$^{-2}$~s$^{-1}$ above 0.85 TeV from the Crab Nebula 
observations). Therefore, it is evident from the above analysis that the blazar Mrk 421 has undergone a high activity state at TeV energies on December 28, 2014. 
\begin{figure}
\begin{center}
\includegraphics[width=0.65\textwidth,angle=-90]{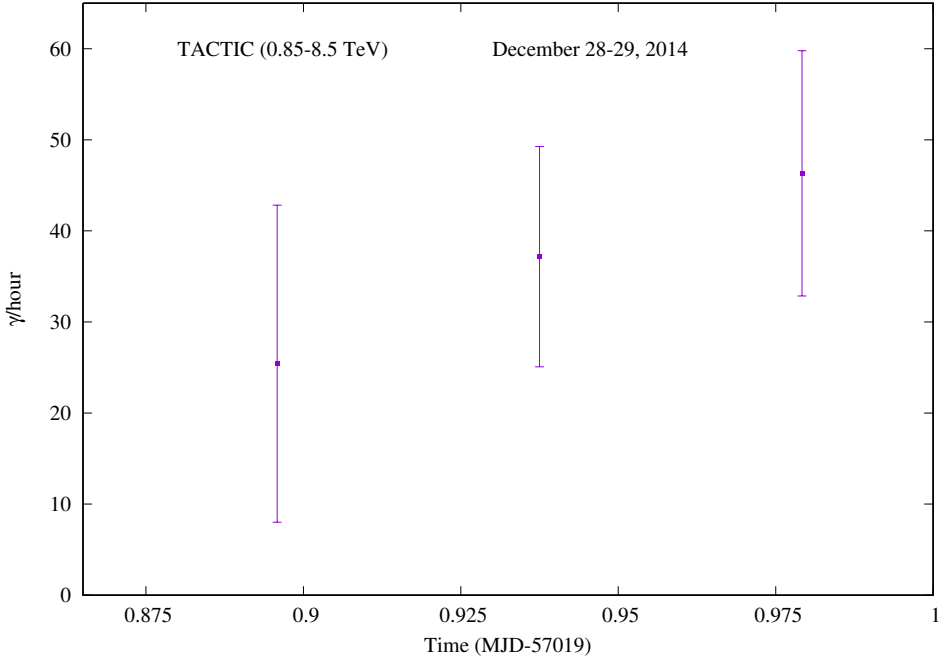}
\caption{Intra-night TeV light curve of Mrk 421 observed with the TACTIC telescope on December 28, 2014.}
\label{fig:Fig2}
\end{center}
\end{figure}

\subsection{Intra-night TeV light curve}
As discussed in Section 3.1, the analysis of TACTIC data results in the statistically significant detection of TeV photons at a rate of (39$\pm$8) $\gamma$- rays 
per hour in a livetime of 2.2 hours. The intra-night TeV light curve of Mrk 421 observed with TACTIC on the night of December 28, 2014 (MJD 57019.89583-57019.97917) is shown in Figure 
\ref{fig:Fig2}. The individual TeV $\gamma$- ray rates in the intra-night light curve correspond to a statistical significance of more than 3$\sigma$. Detailed analysis of the flux measured 
during the flaring activity at smaller time bins will give evidence of very fast variability at TeV $\gamma$- ray energies in Mrk 421. The temporal analysis of such a short duration flare is 
also important because a large fraction of the bolometric luminosity of blazars is produced at GeV-TeV energies. However, the detailed temporal analysis of the intra-night TeV light curve 
does not indicate significant variability because of the large error bars in the $\gamma$- ray rates as shown in Figure \ref{fig:Fig2}. The null hypothesis for constant emission during the 
flaring activity gives a constant TeV $\gamma$- ray rate of (38$\pm$5) $\gamma$- rays per hour corresponding to the reduced-$\chi^2$ and degree of freedom ($\chi_r^2$/dof) value of 0.45/2 with 
the probability $\sim$ 63.5$\%$. This again indicates that the average TeV $\gamma$- ray emission from Mrk 421 during the short duration high activity state observed with TACTIC is compatible 
with the emission level 3-times higher than that of the Crab Nebula.   
\begin{figure}
\begin{center}
\includegraphics[width=1.0\textwidth]{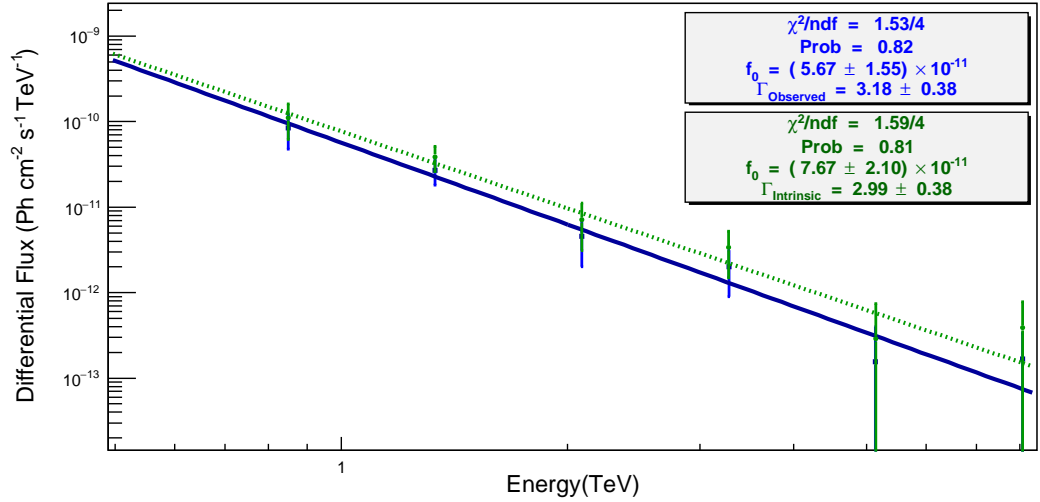}
\caption{Diffrerential energy spectrum of TeV $\gamma$--ray photons detected from Mrk 421 with TACTIC during high activity state on December 28, 2014 (Blue). 
         The corresponding intrinsic VHE spectrum is obtained after correcting for EBL absorption using Dom\'inguez et al. 2011 model [61] (Green).}
\label{fig:Fig3}
\end{center}
\end{figure}

\subsection{$\gamma$- ray differential energy spectrum}
The TACTIC observations of Mrk 421 on the night of December 28, 2014 confirm the detection of statistically significant TeV $\gamma$- ray photons in a short livetime of $\sim$ 2.2 hours.
This enables to obtain the TeV spectrum of the source down to short timescale from the TACTIC observations. The unfolded differential energy spectrum of Mrk 421 derived from 
the TACTIC observations during the high activity state is shown in Figure \ref{fig:Fig3}. The time averaged differential energy spectrum between 0.85-8.5 TeV is described by a simple power 
law of the form :
\begin{equation}
	\frac{dN}{dE} = f_0 \left(\frac{E}{1 TeV}\right)^{-\Gamma}
\end{equation}
with an observed photon spectral index of $\Gamma_{obs}=3.18\pm0.38$ and a normalization constant of $f_0=(5.67\pm1.55)\times10^{-11}$ ph~cm$^{-2}$~s$^{-1}$~TeV$^{-1}$ at 1 TeV.
The mean integral flux above 0.85 TeV is (3.68$\pm$1.07)$\times$ 10$^{-11}$ ph~cm$^{-2}$~s$^{-1}$ which corresponds to $\sim$ 3.2 times the Crab Nebula flux. We have also calculated the 
intrinsic VHE spectrum of the source, considering the absorption of TeV photons by $e^-~e^+$ pair creation due to interaction with the extragalactic background light (EBL) photons. Despite 
being a nearby source at redshift $z=0.031$, a significant absorption of TeV photons is obtained for the EBL model proposed by Dom\'inguez et al. (2011) [61]. The intrinsic VHE spectrum 
corresponding to the above observed spectrum is also described by a power law with photon spectral index $\Gamma_{int}=2.99\pm0.38$ and normalization constant 
$f_0=(7.67\pm2.10)\times10^{-11}$ ph~cm$^{-2}$~s$^{-1}$~TeV$^{-1}$ at 1 TeV. Within error bars the observed and intrinsic spectral indices of Mrk 421 are found to be consitent with the 
previous values reported in the literature [15,16]. The time averaged HE $\gamma$- ray spectrum of the source from \emph{Fermi}-LAT observations on December 28, 2014 is also described by a 
power law with spectral index 1.41$\pm$0.11 in the energy range 100 MeV to 300 GeV and the corresponding integral flux above 100 MeV is found to be 
(1.71$\pm$0.72)$\times$ 10$^{-7}$ ph~cm$^{-2}$~s$^{-1}$.

\subsection{Comparison with near simultaneous HAWC observation}
The High Altitude Water Cherenkov (HAWC) observatory at an altitude of 4.1 km provides continuous survey of the sky and opens a new window for identifying extreme 
VHE flares [62]. The wide field of view and high duty cycle of HAWC provides unique opportunity to observe every TeV source that transits over the observatory 
for up to 6 hours each sidereal day. Long term monitoring of Mrk 421 with the HAWC observatory for 17 months during November 2014-April 2016 gives an average flux of 
(4.53$\pm$0.14)$\times$ 10$^{-12}$ ph~cm$^{-2}$~s$^{-1}$ above 2 TeV under a constant flux emission model [63]. An integral flux of (2.91$\pm$0.38)$\times$ 
10$^{-11}$ ph~cm$^{-2}$~s$^{-1}$ above 2 TeV was measured on December 29, 2014 (MJD 57020.33-57020.58) from $\sim$ 6 hours of observations of Mrk 421 with the HAWC observatory. 
The time averaged integral flux above 0.85 TeV measured with the TACTIC on the night of December 28, 2014 (MJD 57019.89583-57019.97917) is found to be (3.68$\pm$0.64)$\times$ 
10$^{-11}$ ph~cm$^{-2}$~s$^{-1}$ in an observation time of $\sim$ 2.2 hours. The near simultanoeus measurements of the VHE integral flux with TACTIC and HAWC indicate that Mrk 421 
is observed to be in a high activity state at TeV energies. The conversion of integral flux measured with TACTIC from the short term flaring state of Mrk 421 in Crab units above 1 TeV 
also gives a flux value of $\sim$ 3-times the Crab Nebula flux above 1 TeV, which is much higher than the quiescent level (33$\%$ of Crab Nebula flux) of Mrk 421 estimated from the 
long-term light curves for the period 1991-2009 [31].
\section{Results from multi-wavelength observations}

\subsection{Multi-wavelength light curves}
\begin{figure}
\begin{center}
\includegraphics[width=1.0\textwidth]{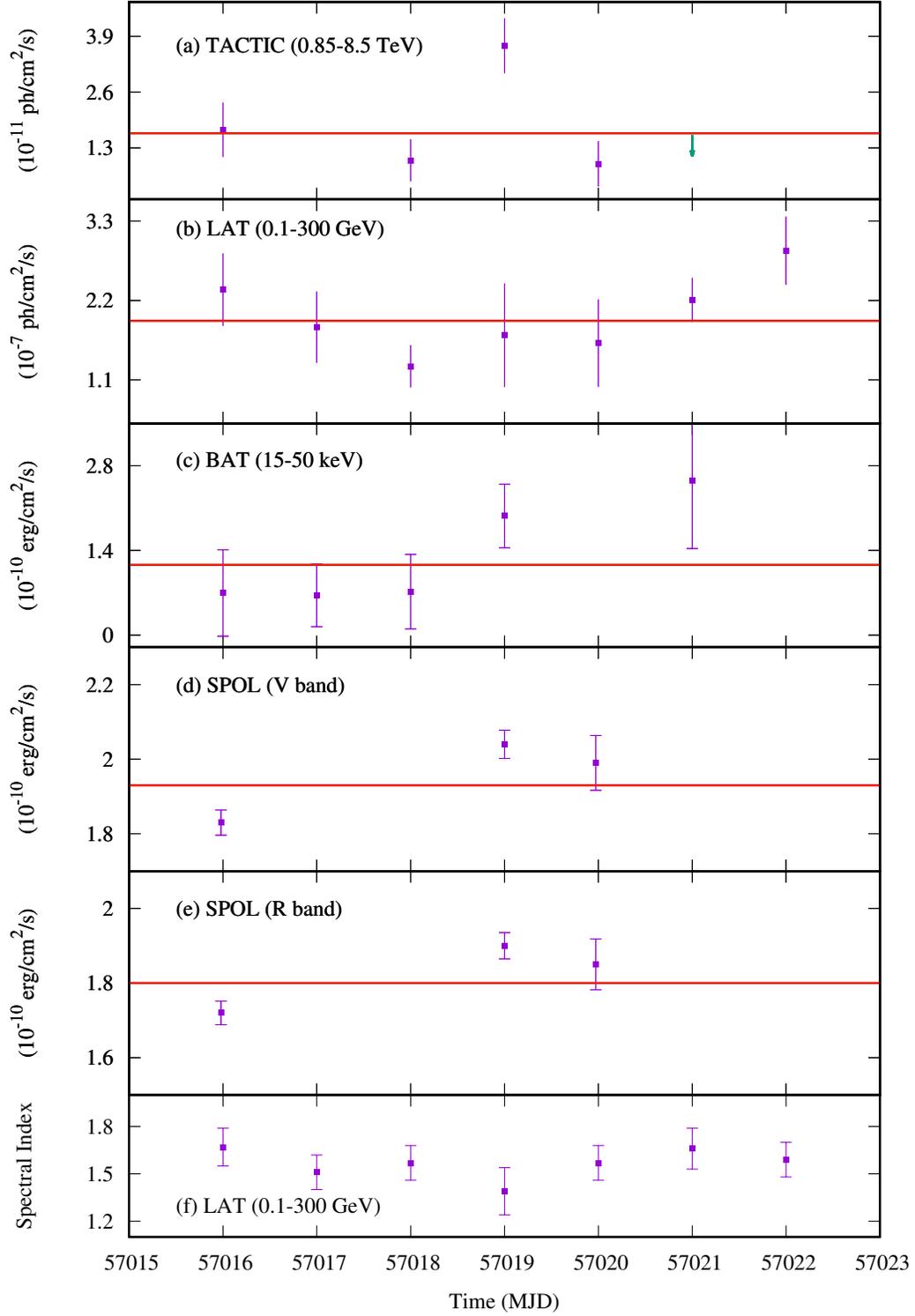}
\caption{Multi-wavelength daily light curves for Mrk 421 during December 25-31, 2014 (MJD 57016-57022). The horizontal solid lines (red) in the panels 
         (a-e) represent average flux level in different energy bands during this period. Bottom panel (f) shows the variation of the HE $\gamma$--ray photon 
         spectral indices measured with \emph{Fermi}-LAT.}
\label{fig:Fig4}
\end{center}
\end{figure}

The multi-wavelength daily light curves of the blazar Mrk 421 during December 25-31, 2014 (MJD 57016-57022) in TeV, GeV, hard X-ray and optical (V and R) bands are shown in 
Figure \ref{fig:Fig4}(a-e). The TeV flux points averaged over one day as shown in Figure \ref{fig:Fig4}(a), represent the integral flux above 0.85 TeV corresponding to the 
TACTIC observations with detection significance more than 2$\sigma$ whereas observations with statistical significance less than 2$\sigma$ are depicted as upper limits. 
We observe that the source undergoes a flaring activity at TeV energies on December 28, 2014 (MJD 57016) with an integral flux of (3.68$\pm$0.64)$\times$ 10$^{-11}$ 
ph~cm$^{-2}$~s$^{-1}$ ($\sim$ 3 times the Crab Nebula flux) above 0.85 TeV, which corresponds to the statistical significance of 5.16 $\sigma$ for TACTIC observation of 
Mrk 421 in a live time of $\sim$ 2.2 hours. The daily averaged light curves shown in  Figure \ref{fig:Fig4}(b-e) are results from the near simultaneous observations of Mrk 421  
with \emph{Fermi}-LAT, \emph{Swift}-BAT and SPOL respectively. The visual inspection of the multi-wavelength light curves shown in Figure \ref{fig:Fig4} indicates that 
TeV emission from the source is variable during this period with a dominant flaring activity observed on December 28, 2014 (MJD 57019). We use  null hypothesis for a constant 
flux emission to characterize the variability present in the emissions from the source in different energy bands. The values of constant flux level and correponding goodness of fit 
obtained from the null hypothesis for the flux points in various energy bands are given in Table \ref{tab:Tab1}. The values of reduced-$\chi^2$ and degree of freedom ($\chi_r^2$/dof) 
along with the probabilities indicate that the flux points measured with TACTIC and SPOL (V and R bands) are not consistent with the constant emission model whereas the HE 
$\gamma$- ray in the energy range 0.1-100 GeV and hard X-ray emission in the energy range 15-50 keV measured with \emph{Fermi}-LAT and \emph{Swift}-BAT respectively are 
constant during this period. The horizontal solid lines (red) in Figure \ref{fig:Fig4}(a-e) represent average flux level in different energy bands. It is evident that the TeV flaring 
activity detected with TACTIC on December 28, 2014 (MJD 57019) is also accompanied by relatively enhanced activity in lower energy bands except \emph{Fermi}-LAT wherein the flux value 
is slightly below the average level. However, the HE $\gamma$--ray emission  measured with \emph{Fermi}-LAT is observed to be gradually increasing after the TeV flaring activity on  
December 28, 2014 (MJD 57019), which may lead to the higher flux level estimated during this period. In Figure \ref{fig:Fig4}(f), we have shown the variation of HE $\gamma$--ray photon 
spectral indices of a power law fit obtained from \emph{Fermi}-LAT daily observations of Mrk 421. No significant trend of spectral hardening with the increasing flux in HE band is 
observed during this period. However, it is to be noted that the HE $\gamma$--ray emission measured from \emph{Fermi}-LAT gradually increases after the relatively high TeV activity state. 
But, due to lack of observations available in other wave-bands during this period, no clear explaination can be given for such activity in the emission from the source. Also, it is beyond 
the scope of this work as we mainly focus on the short term high activity state of Mrk 421 detetcted at TeV energies on December 28, 2014 (MJD 57019) with the TACTIC telescope .        
\begin{table}
\caption{Summary of the constant emission model fit to the flux points reported in the multi-wavelength light curves.}
\begin{center}
\begin{tabular}{ccccc}
\hline
Instrument	&Energy Range		&Average flux						&$\chi_r^2$/dof &Probability\\
\hline
TACTIC		&0.85-8.6 TeV		&(1.64$\pm$0.60)$\times$10$^{-11}$~ph~cm$^{-2}$~s$^{-1}$  &4.56/3	&0.003\\			
LAT		&0.1-300 GeV		&(1.92$\pm$0.21)$\times$10$^{-7}$~ph~cm$^{-2}$~s$^{-1}$   &1.81/6	&0.092\\
BAT		&15-50 keV		&(1.16$\pm$0.33)$\times$10$^{-10}~$erg~cm$^{-2}$~s$^{-1}$ &1.44/4	&0.217\\
SPOL 		&V-band 		&(1.93$\pm$0.07)$\times$10$^{-10}~$erg~cm$^{-2}$~s$^{-1}$ &9.24/2	&0.000\\
SPOL		&R-band 		&(1.80$\pm$0.06)$\times$10$^{-10}~$erg~cm$^{-2}$~s$^{-1}$ &7.40/2	&0.000\\	
\hline
\end{tabular}
\end{center}
\label{tab:Tab1}
\end{table}

\subsection{Variability Analysis}
The multi-wavelength emission from Mrk 421 during December 25-31, 2014 (MJD 57016-57022) is observed to be variable with significant flux changes at TeV energies. 
In order to further quantify the intrinsic source variability, we have estimated fractional variability (F$_{var}$), variability amplitude parameter (A$_{mp}$) and 
relative variability amplitude (RVA) in different energy bands. The variability analysis of multi-wavelength light curves using these methods is possible only when the 
fluctuations in flux points exceed the measurement error. The fractional variability statistic (F$_{var}$) is given by [64] 
\begin{equation}
	F_{var}=\sqrt{\frac{S^2 -E^2}{F^2}}
\end{equation}
and the formal uncertainty in F$_{var}$ is computed by
\begin{equation}
	\Delta F_{var}=\sqrt{\left(\sqrt{\frac{1}{2N}}\frac{E^2}{F^2F_{var}}\right)^2+\left(\sqrt{\frac{E^2}{N}}\frac{1}{F}\right)^2}
\end{equation}  
where $S^2$ is the variance, $E^2$ is the mean square measurement error, $F$ is the mean flux and $N$ is the number of flux points in the light curve.
The estimated values of F$_{var}$ in different energy bands are summarized in Column 2 of Table \ref{tab:Tab2}.  The highest value of F$_{var}$ is obtained 
for TeV observations from the TACTIC whereas measurements with \emph{Fermi}-LAT and \emph{Swift}-BAT in HE $\gamma$- ray and hard X-ray bands respectively 
are error dominated. The values of fractional variability for optical observations are high because of low measurement errors in the flux points. Therefore, 
we observe that the blazar Mrk 421 exhibits significant variability during December 25-31, 2014 at TeV energies. The intrinsic variability of the source in 
soft X-rays can not be probed due to unavailability of observations during this period. 
\par
The percentage variation in the multi-wavelength light curves is calculated through the peak-to-peak variability amplitude parameter (A$_{mp}$) which 
is defined as [65]
\begin{equation}
	A_{mp}=100\times \frac{\sqrt{(F_{max}-F_{min})^2-2\sigma^2}}{F}~~\%
\end{equation}
and the error in A$_{mp}$ is estimated as 
\begin{equation}
	\Delta A_{mp}=100\times \left(\frac{F_{max}-F_{min}}{FA_{mp}}\right)\sqrt{\left(\frac{\Delta F_{max}}{F}\right)^2 + \left(\frac{\Delta F_{min}}{F}\right)^2 
                      + \left(\frac{\Delta F}{F_{max}-F_{min}}\right)^2 A_{mp}^4}~~\%
\end{equation}
where F$_{max}$ and F$_{min}$ are the maximum and minimum flux values with uncertainties $\Delta$F$_{max}$ and $\Delta$F$_{min}$ respectively in each light curve, 
$\Delta$F is the error in mean flux, and $\sigma$ is the average measurement error. The calculated values of A$_{mp}$ for the multi-wavelength light curves is given 
in Column 3 of Table \ref{tab:Tab2}. A maximum peak-to-peak variation is obtained for TeV $\gamma$--rays whereas optical emissions in V and R bands show minimum 
variation during this period. The lowest values of A$_{mp}$ for optical observations are also consistent with the visual inspection of light curves where the peak 
flux values are not clearly identified as compared to TeV light curve. The next parameter for characterizing the variability of the source is the estimation of 
relative variability amplitude (RVA) which we define as [66]
\begin{equation}
	RVA=\frac{F_{max}-F_{min}}{F_{max}+F_{min}}    
\end{equation} 
and the uncertainty on RVA is given by
\begin{equation}
	\Delta RVA=\frac{2}{(F_{max}+F_{min})^2}\sqrt{(F_{max}~\Delta F_{min})^2 + (F_{min}~\Delta F_{max})^2}
\end{equation}  
The RVA values given in Column 4 of Table \ref{tab:Tab2} are also found to be consistent with the general behaviour of blazars wherein maximum variability 
is expected at high energy bands. Above analysis based on the estimation of three variability parameters also indicates the presence of variability in the 
emissions from Mrk 421 in different energy bands with TeV flux points showing maximum variability. 
\begin{table}
\caption{Summary of the results from the intrinsic variability analysis of the multi-wavelength light curves.}
\begin{center}
\begin{tabular}{cccc}
\hline
Energy Range	&F$_{var}$	&A$_{mp}$($\%$)	&RVA\\
\hline
0.85-8.6 TeV	&0.52$\pm$0.17	&95$\pm$35	&0.50$\pm$0.15	\\		
0.1-300 GeV	&0.20$\pm$0.10  &66$\pm$29      &0.38$\pm$0.11	\\
15-50 keV	&0.36$\pm$0.29  &61$\pm$27      &0.59$\pm$0.29	\\
V-band		&0.35$\pm$0.01  &6.2$\pm$2.2    &0.05$\pm$0.01	\\	
R-band 		&0.35$\pm$0.01	&6.9$\pm$2.6    &0.05$\pm$0.01	\\
\hline
\end{tabular}
\end{center}
\label{tab:Tab2}
\end{table}

\subsection{Temporal analysis of TeV light curve}
In order to compute the approximate value of the minimum variability timescale from the TeV light curve shown in Figure \ref{fig:Fig4}(a), we assume that TeV emission
in the flaring activity evolves exponentially during rising or falling of the flux. We define the e-folding timescale as [67]
\begin{equation}
	\tau_{ij} = \left|\frac{\Delta t_{ij}}{ln F_i-ln F_j}\right|
\end{equation}
where $\Delta t_{ij}$ is the time difference between $i^{th}$ and $j^{th}$ flux points i.e.  $F_i$ and $F_j$ respectively in the light curve. The shortest variability timescale 
is calculated as $\tau_v=min(\tau_{ij})$. This method of computing minimum variability timescale uses only two flux measurements and it does not require any fitting of the flux 
points in the light curve. The minimum variability timescale for TeV $\gamma$- ray emission on December 28, 2014 from TACTIC observations of Mrk 421 is found to be 
$\tau_v ~ \sim$ 0.72 days. A conservative estimate of minimum variability timescale in the source frame of Mrk 421 gives $\tau_v/(1+z)=0.69$ days. Such a short variability timescale 
in TeV $\gamma$- ray emission of the source will be useful in constraining the size of emitting region and its location in the jet from the central engine.

\subsection{Spectral Energy Distribution}
\begin{figure}
\begin{center}
\includegraphics[width=0.7\textwidth,angle=-90]{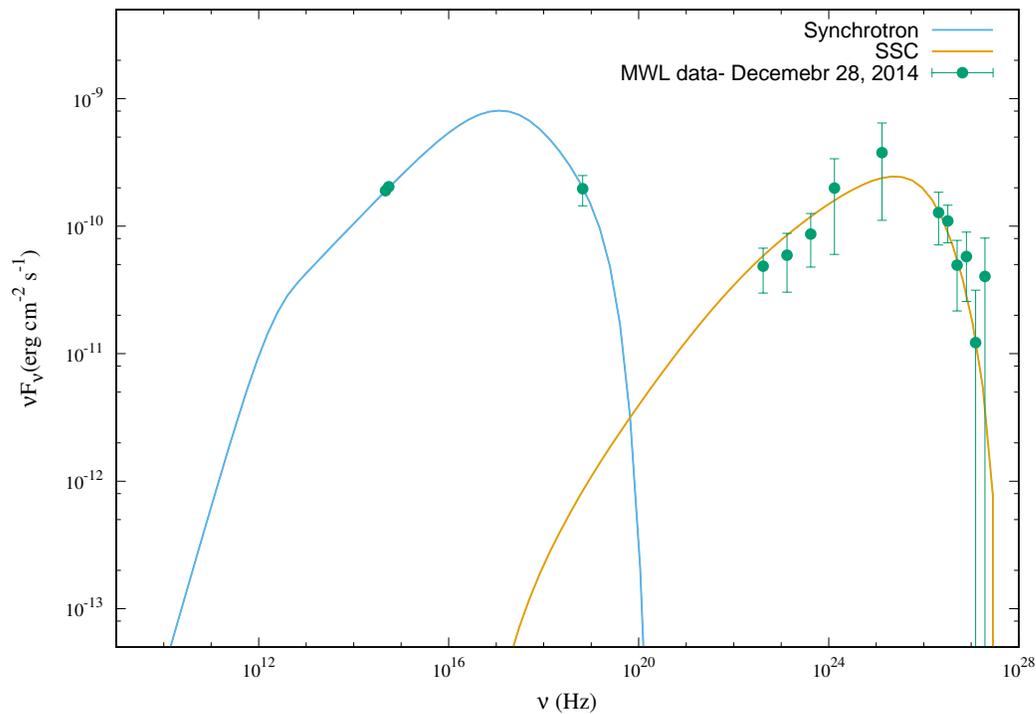}
\caption{Broad-band spectral energy distribution of the blazar Mrk 421 under the frame-work of single zone homogeneous SSC model observed on December 28, 2014. The multi-wavelength 
         data involve near simultaneous observations from SPOL (Optical: V and R bands), \emph{Swift}-BAT (hard X-rays), \emph{Fermi}-LAT (HE $\gamma$- rays) and TACTIC (VHE $\gamma$- rays).
        The TeV flux points correspond to the livetime of 2.2 hours for TACTIC observations on the night of December 28, 2014 whereas flux points from remaining instruments are averaged over 
        1 day. The VHE flux points from TACTIC have also been corrected for EBL absorption using Dom\'inguez et al. (2011) model [61].}
\label{fig:Fig5}
\end{center}
\end{figure}

In order to understand the broad-band emission characteristics of the source during relatively high activity state at TeV energies, we have used a simple one zone leptonic SSC model 
fully described in [68]. In this model, the emission region is assumed to be a spherical blob of radius $R$ moving down the blazar jet with bulk Lorentz factor $\Gamma_j$ at a viewing 
angle $\theta$. The blob is homogeneously filled with a population of relativistic electrons (and positrons) with the differential number density described by a smooth broken power 
law of the form :
\begin{equation}
	n(\gamma)=\frac{K}{\left[ \left(\frac{\gamma}{\gamma_b}\right)^p + \left(\frac{\gamma}{\gamma_b}\right)^q\right]} ; \qquad  \gamma_{min} < \gamma < \gamma_{max} 
\end{equation}
where $\gamma$ is the Lorentz factor of electrons (and positrons) in the blob rest frame, $K$ is the normalization constant, $p$ and $q$ are the low and high energy 
spectral indices before and after the break respectively and $\gamma_b$ is the Lorentz factor corresponding to the break in the electron distribution. $\gamma_{min}$ and 
$\gamma_{max}$ are the Lorentz factors corresponding to the minimum and maximum energies of the electron distribution in the emission region respectively. The 
emitting region is also assumed to be filled with a uniform tangled magnetic field $B$. The relativistic electrons are considered to lose their energy through 
synchrotron process in the magnetic field $B$ and by the inverse Compton scattering off the synchrotron photons produced by them. Due to the relativistic 
bulk motion of the jet, the radiation from emitting region is boosted in the rest frame of the blazar by the Doppler factor given as 
\begin{equation}
	\delta = \frac{1}{\Gamma_j(1-\beta_j cos~\theta)}
\end{equation} 
where $\beta_j$ is bulk speed of jet in the units of speed of light in vacuum, $c$. In case of low viewing angle approximation for blazars, $\delta = \Gamma_j$ corresponding to 
$\theta=1/\Gamma_j$. The size of emission region is constrained by the minimum variability timescale (t$_{var}$) using the relation : 
\begin{equation}
	R \approx \frac{\delta~c~t_{var}}{1+z}
\end{equation}
The broad-band emissions from the blob due to synchrotron and SSC processes are estimated using the convolution of single particle emissivity with the particle 
distribution given in Equation 9. Finally, taking into account the relativistic boosting and other cosmological effects, the flux measured by the observer at energy $E_{obs}$ is 
given by [69]
\begin{equation}
	F (E_{obs}) = \frac{\delta^3 (1+z)}{d^2_L} V j(E)
\end{equation}
where $d_L$ is luminosity distance of the source, V$\left(= 4\pi R^3/3\right)$ is the volume of emission region and $j(E)$ is the emissivity at energy $E$ corresponding to synchrotron 
and SSC processes. We have applied this simple SSC model to reproduce the broad-band emission from Mrk 421 measured on December 28, 2014. The near simultaneous multi-wavelength flux 
measurements along with the SED obtained from the synchrotron and inverse Compton processes are shown in Figure \ref{fig:Fig5}. We observe that the multi-wavelength flux points are broadly 
reproduced by the simple SSC model. The best fit model parameters derived from the broad-band SED modelling in the present work have been summarized in Table \ref{tab:Tab3} and are found to 
be consistent with the values reported in the literature for Mrk 421 [11,16].   
\begin{table}
\caption{Best fit model parameters from the SED fitting of Mrk 421 observations on the night of December 28-29, 2014 using one zone SSC model.}
\begin{center}
\begin{tabular}{lccc}
\hline
Parameter		   		&Symbol      	&Value\\
\hline
Radius of emission region  		&R		&2.11$\times$10$^{16}$ cm\\		
Bulk Lorentz factor of jet 		&$\Gamma_j$ 	&13	\\
Low energy index of particle ditribution &p		&2.21	\\
High energy index of particle ditribution &q            &4.22  \\
Break energy of particle distribution     &E$_b$	&48 GeV \\
Magnetic field                            &B            &0.60 G\\
Particle energy density                   &U$_e$        &2.70$\times$10$^{-3}$ erg~cm$^{-3}$\\
\hline
\end{tabular}
\end{center}
\label{tab:Tab3}
\end{table}
 
\section{Discussion and Conclusions}
We have performed a detailed study of the short term TeV flare of high synchrotron peaked blazar Mrk 421 detected with the TACTIC telescope on the night of December 28, 2014 (MJD 57019) using 
the data collected during December 25-31, 2014 (MJD 57016-57022). The TACTIC telescope has detected 86$\pm$17 TeV $\gamma$--ray photons with a statistical significance of 5.17$\sigma$ in a 
short livetime of 2.2 hours on December 28, 2014. The time averaged VHE $\gamma$--ray rate detected with the TACTIC during this period corresponds to the source activity at the level of $\sim$ 3-times 
the emission from the Crab Nebula. The main focus of this work is to analyze and study this short duration TeV flare of Mrk 421 observed with the TACTIC and near simultaneous activity of the source in 
other wave-bands. The observed and intrinsic differential energy spectra of TeV photons detected with the TACTIC are described by power law with spectral indices 3.18$\pm$0.38 and 2.99$\pm$0.38 respectively 
in the energy range 0.85-8.5 TeV. The corresponding integral flux measured with the TACTIC telescope above 0.85 TeV is obtained to be (3.68$\pm$0.64)$\times$ 10$^{-11}$ ph~cm$^{-2}$~s$^{-1}$. 
The quasi-simultaneous measurements of the integral flux from Mrk 421 with the TACTIC and HAWC observatory characterize the relatively high activity state of the source at TeV energies during 
observations on December 28, 2014.  
\par 
We have also used near simultaneous observations available in other wave-bands from HE $\gamma$- ray to optical observations. From the analysis of the multi-wavelength light curves, it is found that the TeV 
$\gamma$- ray flare observed with the TACTIC on December 28, 2014 is detected without any significant change in lower energy bands. However, this can not be termed as an orphan flare because observations 
of soft X-rays and radio are not available during this period. We have applied the null hypothesis for constant flux to characterize the varaibility of the source in different energy bands in the first step. 
The goodness of fit ($\chi_r^2$/dof and probability) obtained corresponding to the null hypothesis indicates that the emission in TeV band is significantly variable whereas other wave-bands do not show 
significant variability. We have estimated various amplitude parameters in order to further quantify the variability present in the multi-wavelength light curves of Mrk 421 during the period December 25-31, 
2014. We find that the TeV light curve exhibits relatively high values of variability parameters and implies strong variability during the above period. However, the highest value of $RVA=0.59$ is obtained 
for  hard X-rays which implies that the maximum flux is approximately three times the minimum flux in the light curve. But the large fluctuations in the individual flux measurements reduce its intrinsic 
varability with large error bar. The daily HE $\gamma$- ray photon spectral indices also do not show any significant change during this period. The overall behaviour of the intrinsic fractional varaiability 
is found to be consistent with the general trend of high synchrotron peaked blazars where variability amplitude increases with energy [70] and highest variability with F$_{var}\sim$0.52 occurs in TeV flux 
points measured with the TACTIC. This also indicates that the VHE emission originates from a very compact region in the jet and it can be attributed to the change in electron injection or turbulance in the 
jet [71]. The lower values of  variability amplitudes for optical and \emph{Fermi}-LAT observations can be attributed to the fact that variability amplitude is higher at frequencies beyond the synchrotron
and inverse Compton peaks in blazar SED. Also, in high synchrotron peaked blazars like Mrk 421 the optical emission lines are weak and therefore the synchrotron photons 
at X-ray energies are the dominant targets for the inverse Compton scattering to produce the TeV $\gamma$--rays. Therefore, correlated variability at X-ray and TeV energies is expected from the 
single zone leptonic SSC model. The temporal analysis of TeV light curve gives a conservative estimate of the minimum variability timescale of $\sim$ 0.69 days in the source frame. However, the minimum 
variability timescale estimated from the analysis of near simultaneous multi-wavelength light curves with flaring activities will give the strongest possible upper-limit on the size of the active region 
in the jet taking into account the light travel time effects. The data statistics available during this period is not sufficient to perform such detailed temporal analysis of the emission from the source.
\par
The one day broad-band spectral energy distribution of the source using near simultaneous observations on December 28, 2014 can be broadly reproduced by simple one zone leptonic SSC model. 
The model parameters estimated from the best fitting of the SED are found to be in agreement with the values recently reported in the literature for Mrk 421 [72,11,16]. The difference between the 
electron spectral indices $p$ and $q$ is more than the expected value for the pure radiative/synchrotron cooling break in the electron spectrum. This can be attributed to the energy dependent 
acceleration and escape timescales [73,74] which have not been explicitly modelled in the present work. The kinetic energy or power of the jet is estimated from the derived model parameters by 
assuming that the hadrons in the emission region are cold and do not contribute in the radiative process. Under this approximation, the kinetic power of the jet in the source frame is given by 
[75]
\begin{equation}
	P_{jet} \approx \pi R^2 \Gamma^2_j \beta_j c(U_e + U_B + U_p)	
\end{equation}
where $U_e$, $U_B$ are $U_p$ are comoving energy densities corresponding to leptons, magnetic field and cold protons respectively. Using the best fit model parameters given in Table \ref{tab:Tab3}, 
the jet power is estimated to be 7.44$\times$10$^{44}$ erg~s$^{-1}$ which is consistent with the value generally assumed for blazars. The model parameters derived in this work represent one of the 
probable parameter set for Mrk 421, however they may considerably differ from the values estimated using the multi-zone emission models [17] and because of their inherent degeneracy. The fact that 
strictly simultaneous multi-wavelength observations are not available during the TeV flaring activity of Mrk 421 detected with the TACTIC telescope on the night of December 28, 2014, it is difficult 
to provide any firm conclusion about the emission processes involved in the source. However, given that Mrk 421 is known to exhibit frequent flaring activities in all energy bands from radio to TeV 
with short variability timescales, future contemporaneous multi-wavelength observations of short duration flaring activities will help in constraining the parameter space in a relatively better way. 
 
\section*{Acknowledgment}
We thank the anonymous referee for his/her valuable comments which have significantly improved the contents of the manuscript. Authors are thankful to MIRO, PRL, Mt. Abu for providing the 
facility to realuminise the TACTIC mirror facets. We acknowledge the use of public data obtained through Fermi Science Support Center (FSSC) 
provided by NASA and Swift/BAT transient monitor results provided by the Swift/BAT team. Data from the Steward Observatory spectropolarimetric monitoring project were used. 
This program is supported by Fermi Guest Investigator grants NNX08AW56G, NNX09AU10G, NNX12AO93G, and NNX15AU81G. We would also like to thank our colleague Dr. Sunder Sahayanathan for useful 
discussions and suggestions.

\end{document}